\documentclass[a4paper]{iopart}
\usepackage{array,graphicx,float,epsfig,wrapfig}
\usepackage{amssymb}
\usepackage{multirow,booktabs}
\usepackage[nooneline,bf]{caption2}

\begin{document}

\title[Ellipsometry noise spectrum ...]{Ellipsometry noise spectrum, suspension transfer function measurement and closed-loop control of the suspension system in the Q \& A experiment}

\author[S-J Chen, H-H Mei and W-T Ni]{Sheng-Jui Chen$^*$, Hsien-Hao Mei and Wei-Tou Ni}

\address{Center for Gravitation and Cosmology, Department of Physics,\\
National Tsing Hua University,
Hsinchu, Taiwan 30013, Republic of China}

\begin{abstract}
The Q \& A experiment, aiming at the detection of vacuum birefringence predicted by quantum
electrodynamics, consists mainly of a suspended 3.5 m Fabry-Perot cavity, a rotating permanent
dipole magnet and an ellipsometer.  The 2.3 T magnet can rotate up to 10 rev/s, introducing
an ellipticity signal at twice the rotation frequency.  The X-pendulum gives a good isolation ratio
for seismic noise above its main resonant frequency 0.3 Hz.  At present, the
ellipsometry noise decreases with frequency, from $1\times10^{-5}$ rad$\cdot$Hz$^{-1/2}$ at 5 Hz, $2\times10^{-6}$ rad$\cdot$Hz$^{-1/2}$
at 20 Hz to $5\times10^{-7}$ rad$\cdot$Hz$^{-1/2}$ at 40 Hz.  The shape of the noise spectrum
indicates possible improvement can be made by further reducing the movement between the cavity mirrors.
From the preliminary result of yaw motion alignment control, it can be seen that some peaks due to
yaw motion of the cavity mirror was suppressed.  In this paper, we first give a schematic view of
the Q \& A experiment, and then present the measurement of transfer function of the compound
X-pendulum-double pendulum suspension.  A closed-loop control was carried out to verify the
validity of the measured transfer functions.  The ellipsometry noise spectra with and without yaw
alignment control and the newest improvement is presented.
\end{abstract}

\address{$^*$E-mail:  d883374@phys.nthu.edu.tw}



\section{Introduction}
Quantum Electrodynamics (QED) predicts that vacuum is birefringent under the influence of a strong
magnetic field [1-6].  For a $B$ field of 2.5 T, the difference in indices of refraction for
the light with polarization parallel and the light with polarization perpendicular to the $B$ field is
$\Delta n\!\equiv\!n_{\parallel}-n_{\perp}=2.5 \times 10^{-23}$ and proportional to $B^2$. This
tiny effect can be detected by precision determination of the induced ellipticity on
a laser beam \cite{Iacopini}.  The development of
ultrahigh-precision technology in the laser-interferometric gravitational-wave detection community
prompted our thought of its application to this matter \cite{Ni_1}.
\begin{figure}
\begin{center}
\includegraphics[width=\textwidth,height=0.32316\textwidth]{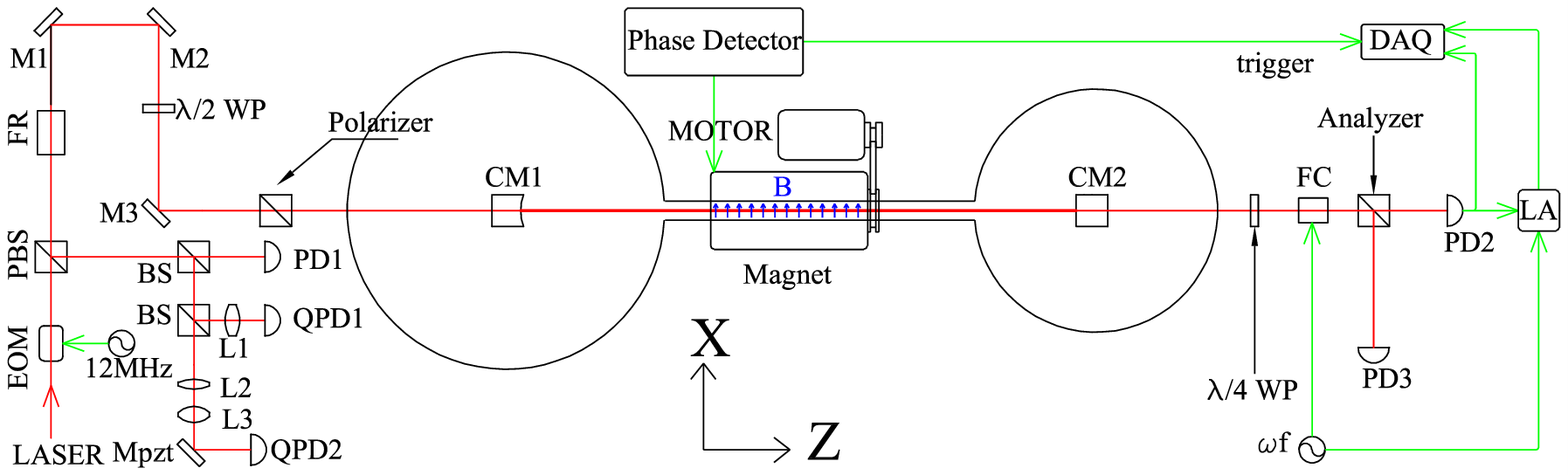}
\caption{Experimental Setup. EOM electro-optical modulator;
$\lambda/2$ WP half-wave plate; $\lambda/4$ WP quarter-wave plate; L1, L2, L3 lenses; M1, M2, M3 steering mirrors;
 PBS polarizing beam splitter; Mpzt Piezo-driven steering mirror; BS beam splitter; FR Faraday rotator; CM1, CM2 cavity mirrors; B magnetic field; PD1, PD2, PD3 photo-receivers; QPD1, QPD2 quadrant photodiodes
; FC Faraday cell; DAQ data acquisition system; LA lock-in amplifier.}
\label{ex_setup}
\end{center}
\end{figure}
First experiment was done in 1993 by Cameron \etal \cite{Cameron} with good upper bounds on vacuum
birefringence and polarization rotation. A pseudo-scalar interaction with electromagnetism
(${\mathcal L}_I\sim{\mathcal \varphi}F_{ij}F_{kl}e^{ijkl}$) was proposed which was empirically allowed in the
study of equivalence principles [8-10]; in terms of Feynman diagram, this interaction gives a
2-photon-pseudo-scalar vertex.  With this interaction, vacuum becomes birefringent and dichroic
[11-13].  In 1994, 3 experiments were put forward and started for measuring the vacuum birefringence:
the PVLAS experiment [14], the Fermilab P-877 experiment [15], and the Q \& A (QED \& Axion) experiment [16];
these experiments were reported in the "Frontier Test of QED and Physics of Vacuum" Meeting in 1998.
Fermilab P-877 experiment was terminated in 2000.  This year in the QED-2005 (QED, Quantum Vacuum
and the Search for New Forces, Les Houches, France, June, 2005) conference, again, 3 experiments
were reported:  the PVLAS experiment [17], the BMV experiment [18], and the Q \& A experiment [19].
A compilation of basic characteristics of these 3 experiments were given in [20].  All 3 experiments
use a high-finesse Fabry-Perot Interferometer (FPI) cavity to enhance the effect to be measured.  The
PVLAS experiment reported a positive measurement of polarization rotation and suggested a possible
interpretation of this result to the existence of a pseudoscalar particle coupled to photons [17, 21].\par
\begin{figure}
\begin{center}
\includegraphics[width=\textwidth,height=0.2729\textwidth]{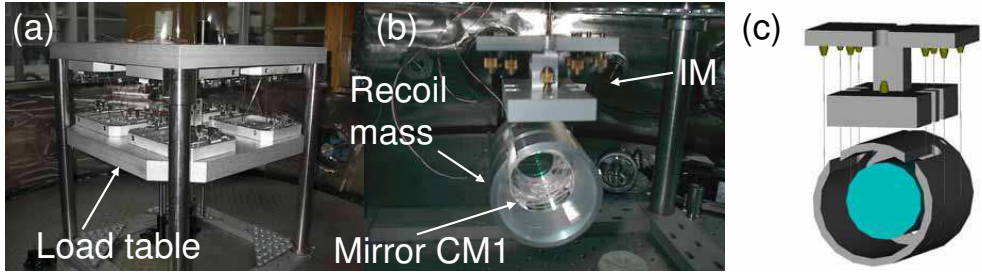}
\caption{(a) Picture of the X-pendulum (b) Picture of the double pendulum (c) CAD drawing of the double pendulum.}
\label{pic_sus}
\end{center}
\end{figure}

In 2002, We had constructed and tested our prototype 3.5 m high-finesse Fabry-Perot interferometer(FPI)
with ellipsometry \cite{Wu}.  Since then, we have been making efforts to improve the stability of
control and the sensitivity of ellipticity detection. Figure \ref{ex_setup} shows the
experimental setup.  The laser beam is phase modulated at 12 MHz for Pound-Drever-Hall locking
technique.  Before entering the FPI, the laser is linearly polarized by a polarizing prism of Glan-Taylor
type with an extinction ratio around 90 dB.  Each mirror of the FPI is suspended by an
X-pendulum \cite{xpen} with a double pendulum as the second stage, as shown in figure \ref{pic_sus}.  X-pendulum was
designed and developed by the TAMA group, a two-dimensional horizontal low frequency vibration isolator. We
followed their design closely and obtained a resonant frequency of 0.3 Hz.
The isolation ratio of the X-pendulum is about 20 dB at 1Hz, 50 dB at 4 Hz, 28 dB at 8 Hz and 34 dB at 20 Hz.
The double pendulum consists of an intermediate mass IM, a recoil mass RM and the mirror CM1.
The longitudinal length of the FPI can be adjusted by applying force to magnets on the CM1 through
coils held by the RM.  A 0.6 m long rotating permanent dipole magnet with a maximum central field
of 2.3 T is located between CM1 and CM2 for producing a polarized vacuum.
The laser is then sent into the FPI to accumulate the ellipticity caused by the vacuum inside the FPI.
After the laser leaving out of the FPI, the acquired ellipticity is transformed into polarization
rotation by a $\lambda/4$ wave-plate.
For lock-in detection, an extra modulation of polarization rotation is applied on the light polarization by a Faraday cell FC \cite{Cameron}.
Finally, the laser is extinguished by another polarizing prism whose transmission axis is aligned
orthogonally to the polarizer's transmission axis.  The transmitted light and reflected light are received by
PD2 and PD3 respectively.  The output of PD3 is used for switching on/off the feedback force on CM1.
PD1, QPD1 and QPD2 are the essential sensors for maintaining the FPI at its working point.  In the
following, we present the transfer function measurement, feedback control and the preliminary results.
Outlook and discussion is given in the end.

\section{Transfer function measurement}

\begin{figure}
\begin{center}
\includegraphics[width=\textwidth, height=0.75\textwidth]{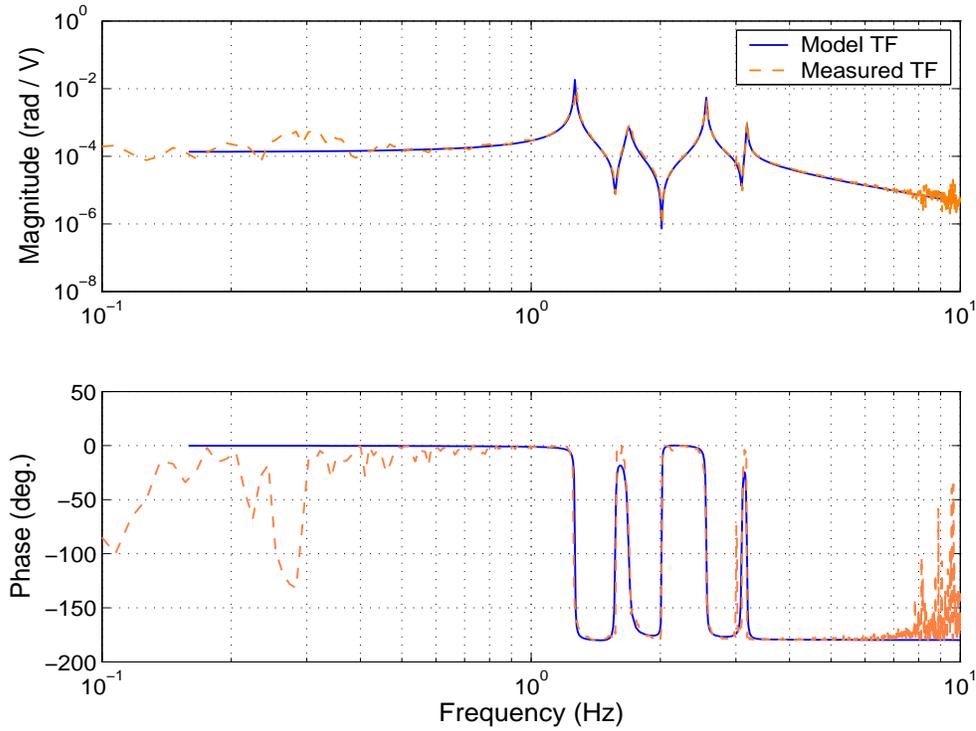}
\caption{The yaw mode (rotation along the vertical axis) transfer function.}
\label{tf_yaw}
\end{center}
\end{figure}

In order to develop length and auto-alignment control servos, we have measured the transfer function (TF) from the
coil-magnet pair actuators to the displacement and rotation of the cavity mirror.
An HP 3-axis heterodyne interferometer capable of sensing two rotational and one longitudinal
motions was used to detect the movement of the cavity mirror. The HP 3-axis interferometer was
mounted on the optical bench in which the X-pendulum is seated.
The cavity mirror CM1, the intermediate mass IM and the recoil mass RM are arranged in a double
pendulum configuration and suspended from the load table of the X-pendulum, as shown in figure \ref{pic_sus}.
The transfer function in the longitudinal d.o.f. is simply the transfer function of a simple pendulum.
For transfer funtion in rotational(yaw) d.o.f., it is more complicated because of the moment-of-inertias of
CM1 and RM are not matched in this design.
The measured yaw mode transfer funciton and the computer model are shown in figure \ref{tf_yaw}.
A active-damping control was carried out to verified the measured transfer functions.
The outputs of the HP interferometer were taken as the error signals in the control loop.
The loop filters were designed and optimized by the closed-loop simulation with the model transfer functions,
and realized by a digital control unit.
The filter for yaw mode control basically consists of an integrator and a bi-quadratic filter (a complex zero and a complex pole),
in order to provide enough DC gain and phase lead in frequency range from 1 Hz to 4 Hz.
In this control test, the rms amplitude was suppressed
from 4.05 $\mu\mbox{m}$ to 0.23 $\mu\mbox{m}$ for the longitudinal motion and from
15.5 $\mu\mbox{rad}$ to 4.4 $\mu\mbox{rad}$ for the yaw motion \cite{QNA_qed2005}.  A similar active damping control
with more axes damped and faster loop rate is to be added in the main ellipsometry measurement.

\section{Feedback control}
\subsection{Longitudinal control}
Because of the large resonant displacement of the X-pendulum, the frequency range of the laser
is not sufficient to track the cavity resonant frequency.  Thus, the error signal for longitudinal control
has to be fed back into two paths:  one for frequency of the laser with a dynamic range of
$\pm$ 120 MHz and a bandwidth over 100 kHz;  another one for cavity length control with a
dynamic range of $\pm3$ mm($\sim\pm240$ GHz) and a bandwidth of about 200 Hz.  In the first path, a typical loop filter consisting of an
integrator and a lag compensator is used.  In the second path, the loop filter is a first-order
low pass filter whose cut-off frequency is 100 Hz in series with a lag compensator.  The ratio of
DC gains of these two paths determines the stability of the closed-loop control.
The unit gain frequency in this control is about 13.3 kHz.
\subsection{Alignment control}
\begin{figure}
\begin{center}
\includegraphics[width=\textwidth, height=0.75\textwidth]{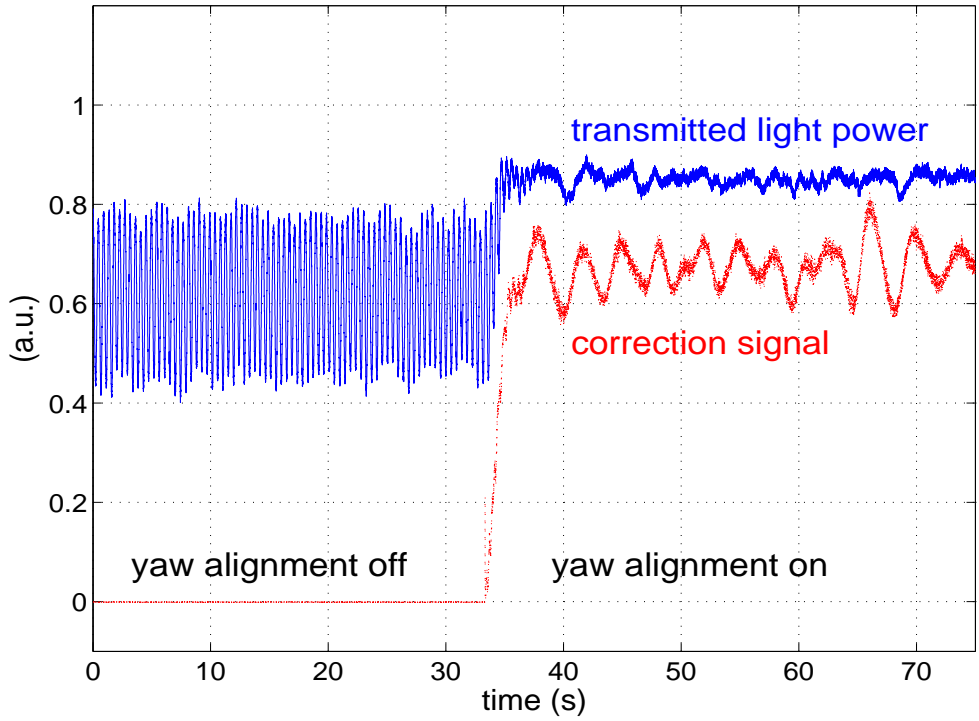}
\caption{The lower curve is the correction signal on the yaw motion actuator.  When the
control is on, the light power transmitted by the FPI increased and its fluctuation decreased.}
\label{tr_alignment}
\end{center}
\end{figure}

The differential wavefront sensing technique \cite{Morrison} is adopted here.  We use a quadrant photodiode (QPD)
with a resonant circuit \cite{Heinzel} as our wavefront sensor.
An active control on centering the interference pattern on the QPD was employed.
The bandwidth for the yaw alignment control is about 15 Hz.
The loop filter is the same as that used in the active-damping test.
The transient response while closing the yaw alignment control loop is plotted in figure \ref{tr_alignment}.
The effect of the yaw alignment control can also be seen in the ellipticity noise spectrum in figure \ref{noise_floor}.
With yaw alignment control, the residual angle fluctuation of the cavity mirror is about 4
$\mu\mbox{rad}$  and will be further reduced by optimizing the control servo.
Now, only one out of four degrees of freedom is actively controlled.
The three remaining d.o.f.'s will be controlled soon after this conference.

\section{Preliminary results}
\begin{figure}
\begin{center}
\includegraphics[width=\textwidth, height=0.66125\textwidth]{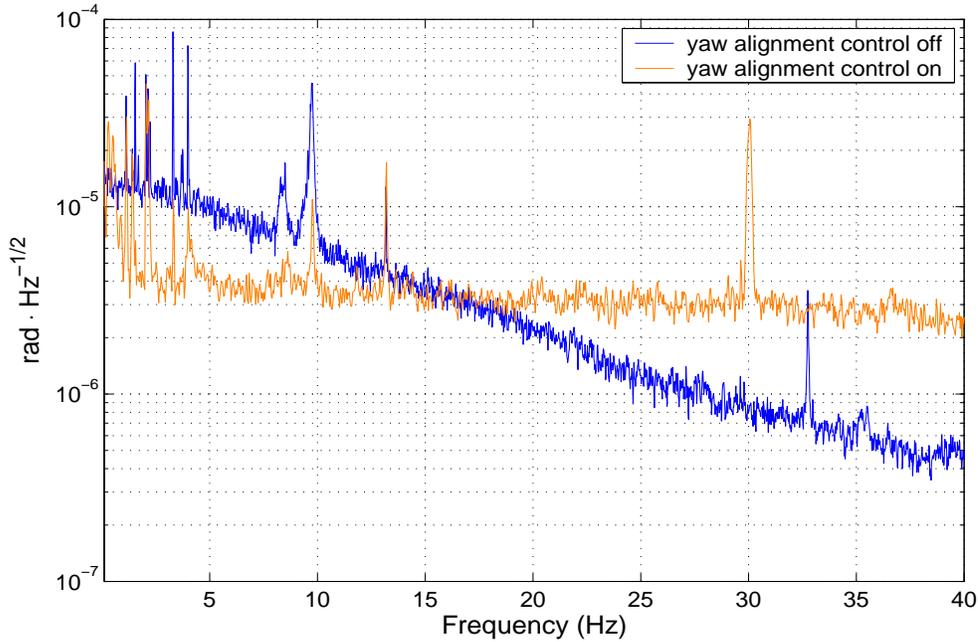}
\caption{Ellipticity noise spectrum.  The detection band is from 10 Hz to 20 Hz.}
\label{noise_floor}
\end{center}
\end{figure}

The light intensity $I$ received by PD2 in figure \ref{ex_setup} can be
described by Malus' law:
\begin{equation}
I=I_0\{\sigma^2+[\eta_0\cos(\omega_ft)+\Psi(t)+\alpha(t)]^2\}\,,
\end{equation}
where $I_0$ is
the incident light intensity on the analyzer, $\sigma^2$ is the extinction ratio of the analyzer,
$\eta_0\cos\omega_ft$ is the rotation modulation of polarization with the modulation angular
frequency $\omega_f$ and modulation depth $\eta_0$.  $\Psi$ is the ellipticity acquired during
round trips inside the magnetic field region of length $L_B$ and is written as
\begin{equation}
\Psi=\frac{2F}{\pi}\frac{\Delta\phi}{2}\sin(2\theta)=\frac{2F}{\pi}\frac{1}{2}\left(\frac{2\pi}{\lambda}L_B\Delta \label{I_extinction}
n\right)\sin(2\omega_mt)\,,
\end{equation}
where $2F/\pi$ is the enhancing factor provided by the FPI with finesse $F$, $\lambda$ is the light wavelength, $\theta$ is the angle
between the magnetic field and the polarization of light, and $\omega_m$ is the rotating angular
frequency of magnet.  The term $\alpha(t)$ is the additional ellipticity and rotation.  It represents
the noise of the apparatus which may be contributed by, for example, the anisotropies of
mirrors' coatings \cite{Jacob}, the movement of cavity mirror and the movement of beam spot on the
analyzer \cite{Cameron}, etc.
We set $\theta(t)=\pi/2$ during runs for measuring the noise floor of the apparatus.
After demodulation of $I$, the term $\alpha(t)$ is obtained.
Figure \ref{noise_floor} shows the spectrum of $\alpha(t)$,
the spectral density decreases with frequency,
from $1\times10^{-5}$ rad$\cdot$Hz$^{-1/2}$ at 5 Hz, $2\times10^{-6}$ rad$\cdot$Hz$^{-1/2}$ at 20 Hz
to $5\times10^{-7}$ rad$\cdot$Hz$^{-1/2}$ at 40 Hz with sensitivity goal in this second phase $5\times 10^{-8}\;\mbox{rad}\cdot\mbox{Hz}^{-1/2}$.
One possible reason for this shape of spectrum is
the displacement of the suspended cavity mirror whose spectral density also decreases with frequency.
The movement of the cavity mirror results in the fluctuation of the light intensity $I$
for two reasons.  One is the change of relative angle between the birefringence axis of
cavity mirror and the light polarization.  The other is the
cavity mirror's phase anisotropy being different for different spot of the cavity mirror.
As can be seen from figure \ref{noise_floor}, the spectral density below 15 Hz is lowered by the alignment
control.  This indicates possible improvement can be made by further reducing the movement between
cavity mirrors.  Above 15 Hz, extra noise is introduced by the alignment control loop.
We are still working on the noise reduction, to find out where the noise is from and to eliminate it.

\section{Outlook and discussion}
In the near future, things to be done include:  (i)Add alignment control in the rest 3 d.o.f.'s and refine the control;
(ii)Increase the finesse of the FPI;  (iii)Add a polarization maintaining fiber as a simple mode cleaner \cite{Mei}.
Hopefully we can achieved our goal sensitivity of $5\times10^{-8}$\;\mbox{rad}$\cdot \mbox{Hz}^{-1/2}$ at 10$\sim$20 Hz.
For the next phase (third phase) after this, with a 5-fold improvement on optical sensitivity, a 5 m
rotating permanent magnet, and a FPI with length extended to 7 m, vacuum birefringence would be
in our reach \cite{chen}.  In this second phase,
we will be able to perform an independent measurement on the polarization rotation to compare with
the PVLAS result [17, 21].
\par
\quad \par
We would like to thank the National Science Council for supporting this work (NSC 93-2112-M-007-022,
NSC 94-2112-M-007-012).


\section*{References}
\begin{thebibliography}{40}
\bibitem{Halpern} Halpern O. 1933 {\it Phys. Rev.} {\bf 44} 855
\bibitem{Euler} Euler E. 1936, Ann der Phys. (Leipzig), {\bf 26} 398 ; Heisenberg W. and Euler E. 1936 Zeits. Fur Phys. {\bf 98} 714
\bibitem{Bialincka} Bialynicka-Birula Z. and Bialynicki-Birula I. 1970 {\it Phys. Rev.} D {\bf 2}  2341
\bibitem{Adler} Adler S. L. 1971 {\it Ann. Phys. (USA)} {\bf 87} 599
\bibitem{Iacopini} Iacopini E and Zavattini E 1979 {\it Phys. Lett.} B {\bf 85} 151
\bibitem{Ni_1} Ni W-T, Tsubono K, Mio N, Narihara K, Chen S-C, King S-K and Pan S-S 1991 {\it Mod. Phys. Lett.} A {\bf 6} 3671
\bibitem{Cameron} Cameron R \etal 1993 {\it Phys. Rev.} D {\bf 47} 3707
\bibitem{Ni_2} Ni W-T 1974 {\it Bull. Amer. Phys. Soc.} {\bf 19} 655
\bibitem{Ni_3} Ni W-T 1977 {\it Phys. Rev. Lett.} {\bf 38} 301
\bibitem{Ni_4} Ni W-T 1973 A Nonmetric Theory of Gravity (Montana State University, Bozeman, Montana,
USA), preprint.  The paper is available via http://gravity5.phys.nthu.edu.tw/webpage/article4/index.html.
\bibitem{ref_4_pvlas} Sikivie P. 1983 {\it Phys. Rev. Lett.} {\bf 51} 1415;  Anselm A. A. 1985 Yad. Fiz. {\bf 42} 1480;
Gasperini M. 1987 {\it Phys. Rev. Lett.} {\bf 59} 396
\bibitem{ref_5_pvlas} Maiani L., Petronzio R. and Zavattini E. 1986 {\it Phys. Lett.} B {\bf 175} 359
\bibitem{ref_6_pvlas} Raffelt G. and Stodolsky L. 1988 {\it Phys. Rev.} D {\bf 37} 1237
\bibitem{Pengo} Pengo R. \etal 1998 Frontier Test of QED and Physics of the Vacuum 59, ed. E Zavattini \etal (Sofia: Heron Press); and references there in.
\bibitem{Nezrick} Nezrick F. 1998 Frontier Tests of QED and Physics of Vacuum 71, ed. E Zavattini \etal (Sofia: Heron Press); and reference there in.
\bibitem{Ni_5} Ni W-T 1998 Frontier Tests of QED and Physics of Vacuum 83, ed. E Zavattini \etal (Sofia: Heron Press); and reference there in.
\bibitem{pvlas_qed2005} Cantatore G. \etal talk on QED-2005 http://arachne.spectro.jussieu.fr/QED2005/Talks/Cantatore.pdf
\bibitem{BMV_qed2005} Rizzo C. \etal talk on QED-2005 http://arachne.spectro.jussieu.fr/QED2005/Talks/Robilliard\_Rizzo.pdf
\bibitem{QNA_qed2005} Chen S-J \etal talk on QED-2005 http://arachne.spectro.jussieu.fr/QED2005/Talks/Chen.pdf
\bibitem{chen} Chen S-J, Mei S-H and Ni W-T 2003 {\it Preprint} hep-ex/0308071
\bibitem{Cantatore} Cantatore G. \etal 2005 {\it Preprint} hep-ex/0507107
\bibitem{Wu} Wu J-S, Ni W-T and Chen S-J 2004 {\it Class. Quant. Grav.} {\bf 21} S1259 ({\it Preprint} physics/0308080)
\bibitem{xpen} Tatsumi D, Barton M A, Uchiyama T and Kuroda K 1999 \RSI {\bf 70} 1561; and references therein
\bibitem{Morrison} Morrison E, Beers B J, Robertson D I and Ward H 1994 {\it Appl. Opt.} {\bf 33} 5037, 5041
\bibitem{Heinzel} Heinzel G. Ph.D. Thesis, University of Hannover 1999; MPQ Report 243 Feb. 1999
\bibitem{Mei} Mei H-H \etal 2005 presented in the 6th Edoardo Amaldi Conference on Gravitational Waves, June, 2005 (Okinawa,
Japan) ({\it Preprint} physics/0508153)
\bibitem{Jacob} Jacob D, Vallet M, Bretenaker F, Le Floch A and Oger M 1995 {\it Opt. Lett.} {\bf 20} 671
\end{thebibliography}
\end{document}